\documentclass[12pt]{iopart}

\usepackage{iopams}

\usepackage{graphicx}

\begin{document}

\title[Entanglement manipulation for mixed   ... ]
{Entanglement manipulation for mixed states in a multilevel atom
interacting with a cavity field }

\author[M. Abdel-Aty et al.]{M. Abdel-Aty$^{*,}$\footnote{%
E-mail: abdelatyquant@yahoo.co.uk}, M. R. B. Wahiddin$^{*}$ and A.-S. F.
Obada$^{**}$
}

\address{$^*$Centre for Computational and Theoretical Sciences, Kulliyyah of
Science, Malaysia, 53100 Kuala Lumpur, Malaysia
\\
$^{**}$Mathematics Department, Faculty of Science, Al-Azhar
University, Nasr City, Cairo, Egypt
}

\begin{abstract}
We derive an explicit formula for an entanglement measure of mixed
quantum states in a multi-level atom interacting with a cavity field
within the framework of the quantum mutual entropy. We describe its
theoretical basis and discuss its practical relevance (especially in
comparison with already known pure state results). The effect of the
number of levels involved on the entanglement is demonstrated via
examples of three-, four- and five-level atom. Numerical
calculations under current experimental conditions are performed and
it is found that the number of levels present changes the general
features of entanglement dramatically.
\end{abstract}

\pacs{32.80.-t, 42.50.Ct, 03.65.Ud, 03.65.Yz}


\maketitle

\section{Overview}

Compared to \textrm{the long history of the theoretical understanding of
entanglement of atom-field systems extending over many decades [1,2],
intensive experimental investigations started only recently involving
different systems [3]. }Entanglement lies at the heart of quantum mechanics,
and is profoundly important in quantum information. To identify the
fundamentally inequivalent ways quantum systems can be entangled is a major
goal of quantum information theory [3]. It might be thought that there is
nothing new to be said about bipartite entanglement if the shared state is
pure, but in a recent paper [4] it has been shown that exact coherence of
the atom is in general never regained for a two-level model with a general
initial pure quantum state of the radiation field. Also, it has been shown
that the purification of the atomic state is actually independent of the
nature of the initial pure state of the radiation field.

From the viewpoint of the Phoenix-Knight [5] entropy formalism, the quantum
field entropy and entanglement of a coherent field interacting with a
three-level systems have been investigated [6]. However, the method used in
those papers cannot be applied when the system is taken to be initially in a
mixed state. A method using quantum mutual entropy to measure the degree of
entanglement in the time development of the two-level system model has been
adopted in {[7]}. The question of how mixed a two-level system and a field
mode may be such that free entanglement arises in the course of the time
evolution according to a Jaynes-Cummings type interaction has been
considered [8-11].

It is important to point out that further insights into the dynamics of the
multi-level systems may be helpful in developing quantum information theory {%
[12]}. Recently, there is much interest in multi-level quantum systems to
represent information [13-15]. It was demonstrated that key distributions
based on multi-level quantum systems are more secure against eavesdropping
than those based on two-level systems [13]. Key distribution protocols based
on entangled three-level systems were also proposed {[16]}. The security of
these protocols is related to the violation of the Bell inequality. The
multi-level system provides in this context a much smaller level of noise {%
[17,18].} Rydberg atoms crossing superconductive cavities are an almost
ideal system to generate entangled states, and to perform small scale
quantum information processing {[19]}. In this context entanglement
generation of multi-level quantum systems was also reported {[12,20-23].}

\textrm{Our motivation is to suggest the use of quasi-mutual entropy
for measuring quantum entanglement in the multi-level systems
(further elaborated below). This is because the quasi-mutual entropy
can be thought of as the original entanglement measure of mixed
(rather than simply pure) input states. } Motivated by recent
experiments we analyze the entanglement degree of a multi-level
system. Using an appropriate representation and without using the
diagonal approximation, an explicit expression for a mixed state
entanglement is derived. \textrm{Although various special aspects of
mixed states entanglement have been investigated previously, the
general features of the dynamics, when a multi-level system is
considered, have not been treated before and the present paper
therefore fills a gap in the literature. }In the present work we
consider the situation for which the multi-level system is initially
in a mixed state. We essentially generalize the entanglement degree
due to the quasi-mutual entropy, usually employed in the two-level
system, to the multi-level system interacting with a cavity field.
The physical situation which we shall refer to, belongs to the
experimental domains of cavity quantum electrodynamics.

\bigskip

\bigskip \textrm{The plan of the remainder of the paper as follows. In
Section 2, we go through a more rigorous set of definitions leading up to
the exact solution of the multi-level system a}nd give exact expression for
the unitary operator $U_{t}$ involved\textrm{. In Section 3, we consider one
of the intermediate definitions namely the accessible entanglement degree
and develop several results related to this quantity. These include a more
convenient expression that automatically takes into account an arbitrarily
number of atomic levels, where the atom is initially in the mixed state,
depending on both the value of the accessible entanglement and on the
measurements required for its definition. In Section 4, we apply the theory
to study a few examples in detail. In particular, we show how difficult it
is to derive rigorously the entanglement involving more than three-levels.
The final part of this article is devoted to some important developments of
entanglement measures and we close the paper with a list of open questions.}

\section{The multi-level system}

We start by devoting this section to a brief discussion on\textrm{\
the multi-level atom [24,25] }being it the model describing the
interaction between a single multi-level atom and a quantized cavity
field. To set the stage, we first begin by describing the
multilevel-atom model. Therefore, the physical system on which we
focus is an $m$-level. The atom interacts with a high Q-cavity which
sustain a number of modes of the field with frequencies {$\Omega_j,
j=1,2,...,m-1$}. We denote by $\hat{a}_{j}$ and
$\hat{a}_{j}^{\dagger }$ the annihilation and creation operators for
the field mode j, and $\omega _{j}$ is the frequency associated with
the level of the atom. We assume that the mode i affects the
transition between the upper atomic level and the level (i+1).
Therefore in the rotating wave approximation we can cast the
Hamiltonian of the system in the form [24] ($\hbar =1)$
\begin{equation}
\hat{H}=\hat{H}_{0}+\hat{H}_{1},  \label{1}
\end{equation}
where the Hamiltonian for the interacting system $\hat H_0$ is
given by
\begin{eqnarray}
\hat{H}_{0} &=&\sum_{j=1}^{m-1}\Omega _{j}\hat{a}_{j}^{\dagger }%
\hat{a}_{j}+\sum_{i=1,2,..m} \omega _{i}\left| i\right\rangle
\left\langle i\right|.
\end{eqnarray}
The interaction Hamiltonian between the atomic system and the
cavity field is given by
\begin{eqnarray}
\hat{H}_{1} =\sum_{j=1}^{m-1}(\lambda _{j}(\hat S_{1,j+1}\hat
a_j+h.c.). \label{2}
\end{eqnarray}
The transition in the $m$-level atom is characterized by the
coupling $\lambda_{i}$. The
operator $\widehat{S}_{ii}$ describes the atomic population of level $%
|i\rangle _{A}$ with energy $\omega _{i},(i=1,2,...,m)$ and the operator $%
\widehat{S}_{ij}=|i\rangle\langle j|, (i\neq j)$ describes the
transition from level $|i\rangle _{A}$ to level $|j\rangle _{A}$.

We have applied the rotating wave approximation discarding the
rapidly oscillating terms and selecting the terms that oscillate
with minimum frequency \textrm{[26]}. The resulting effective
Hamiltonian may be written as
\begin{eqnarray}
\hat{H}_{0} &=&\left( \omega _{1}-\Delta \right)
I+\sum\limits_{j=1}^{m-1}\Omega _{j}\left( \hat{a}_{j}^{\dagger }\hat{a}_{j}-%
\hat{S}_{j+1,j+1}\right) ,
\\
\hat{H}_{1} &=&\Delta \hat{S}_{11}+\sum_{j=1}^{m-1}\lambda
_{i}\left(
\hat{S}_{1,j+1}\hat{a}_{i}+\hat{S}_{j+1,1}\hat{a}_{i}^{\dagger
}\right) .
\end{eqnarray}
We have used $\sum_{i=1}^m=I$. Here we assume that the detuning
parameter $\Delta $ is given by
\[
\Delta =\omega_1-\omega_{j+1}-\Omega _{j},\quad j=1,2,.... m-1.
\]
It can be shown that $\hat{H}_{0}$ and $\hat{H}_{1}$ are constants
of motion,
\begin{equation}
\lbrack \hat{H}_{0},\hat{H}_{1}]=[\hat{H},\hat{H}_{0}]=0.
\end{equation}
We assume that, before entering the cavity, the atom is prepared in a mixed
state. Mixed states arise when there is some ignorance with respect to the
system, so that consideration has to be given to the possibility that the
system is in any one of several possible states, $S_{ii}$, each with some
probability, $\gamma _{i}$, of being realized. To this end, the initial
state of the atom can be written in the following form
\begin{equation}
\rho =\left( \gamma _{1}\widehat{S}_{11}+\gamma _{2}\widehat{S}_{22}+\gamma
_{3}\widehat{S}_{33}+.............+\gamma _{m}\widehat{S}_{mm}\right) \in
S_{A},
\end{equation}
where $\gamma _{i}\geq 0,$ and $\sum\limits_{i=1}^{m}\gamma _{i}=1.$ In
terms of quantum information processes, an understanding of mixed states is
essential, as it is almost inevitable that the ideal pure states will
interact with the environment at some stage.

Also we suppose that the initial state of the field is given by
\begin{equation}
|\varpi_1\rangle =\left( \sum_{n_{1},n_{2}.....=0}^{\infty
}b_{n_{1}}b_{n_{2}}.....b_{n_{m-1}}|n_{1},n_{2}.....n_{m-1}\rangle\right)
\in {S}_{F},
\end{equation}
where $b_{n_{i}}=\langle \varpi |n_{i}\rangle$, $b_{n_{i}}^2$
being the probability distribution of photon number for the
initial state. The continuous map $\mathcal{E}_{t}^{\ast }$
describing the time evolution between the atom and the field is
defined by the unitary operator generated by $\hat{H}$ such that
\begin{eqnarray}
\mathcal{E}_{t}^{\ast } &:&{S}_{A}\longrightarrow {S}_{A}\otimes {\ S}_{F},
\nonumber \\
\mathcal{E}_{t}^{\ast }\rho &=&\hat{U}_{t}\left( \rho \otimes \varpi \right)
\hat{U}_{t}^{\ast }, \\
\hat{U}_{t} &\equiv &\exp \left( -\frac{i}{\hbar }\int \hat{H}(t)dt\right) .
\nonumber
\end{eqnarray}
where $\varpi=|\varpi_1\rangle\langle\varpi_1|$. Bearing these
facts in mind we find that the evolution operator $\hat{U}_{t}$
takes the next from
\begin{equation}
\hat{U}_{t}\equiv \exp \left( -\left( \omega _{1}-\Delta \right) t\right) %
\left[ \prod_{j=1}^{m-1}\exp \left( -i\Omega _{j}\widehat{N}_{j}t\right) %
\right] \exp \left( -i\int_{0}^{t}\hat{H}_{1}dt\right) .
\end{equation}
where $\widehat{N}_{j}=\hat{a}_{j}^{\dagger
}\hat{a}_{j}-S_{j+1,j+1}.$ The first two factors in equation (11)
produce phases that will not affect the results that follow, while
calculations of the third factor show that it takes the following
compact matrix form

\begin{equation}
\exp \left( -i\hat{H}_{1}t\right) =\exp \left( -\frac{i}{2}\Delta t\right) %
\left[
\begin{array}{c}
\widehat{U}_{0} \\
\widehat{U}_{1}^*
\end{array}
\begin{array}{c}
\widehat{U}_{1} \\
\widehat{U}_{2}
\end{array}
\right] ,
\end{equation}
where $\widehat{U}_{0}$ is the single element matrix
$\{\widehat{U}_{1}\}$ which takes the following form
\begin{equation}
\widehat{U}_{11}=\cos \widehat{\mu }_{n}t-\frac{i\Delta }{2}\frac{\sin
\widehat{\mu }_{n}t}{\widehat{\mu }_{n}}.
\end{equation}
The matrix $\widehat{U}_{1}^*$ is the $1\times (m-1)$ row matrix $\{\widehat{U%
}_{1k}\},$ where
\begin{equation}
\widehat{U}_{1k}=-i\frac{\sin \widehat{\mu }_{n}t}{\widehat{\mu }_{n}}%
\lambda _{k}\widehat{a}_{k},\qquad k\in \{1,2,3,......,m-1\}
\end{equation}
and $U_{BA}$ its Hermitian conjugate. Finally the matrix
$\widehat{U}_{2}$ is the $(m-1)\times (m-1)$ square matrix
$\{\widehat{U}_{ij}\}$ of which the elements can be written as
\begin{equation}
\widehat{U}_{ij}=\delta _{ij}\exp \left( -\frac{i}{2}\Delta t\right)
-\lambda _{i}\hat{a}_{i}^{\dagger }v^{-1}\left( \cos \widehat{\mu }%
_{n}t-\exp \left( -\frac{i}{2}\Delta t\right) +\frac{i\Delta }{2}\frac{\sin
\widehat{\mu }_{n}t}{\widehat{\mu }_{n}}\right) \lambda _{j}\hat{a}_{j},
\end{equation}
with $i,j=1,2,.....,m-1$ and
\begin{equation}
\widehat{\mu }_{n}=\left( \frac{\Delta ^{2}}{4}+\sum\limits_{i=1}^{m-1}%
\lambda _{i}^{2}\widehat{a_{i}}\hat{a}_{i}^{\dagger }\right) ^{\frac{1}{2}%
},\qquad v^{-1}=\sum\limits_{i=1}^{m-1}\lambda _{i}^{2}\widehat{a_{i}}\hat{a}%
_{i}^{\dagger }\quad
\end{equation}
Having obtained the explicit form of the unitary operator $U_{t}$,
we are therefore able to discuss the entanglement of the system.

\section{Derivation of the entanglement degree}

Entanglement is recognized nowadays as a key ingredient for fundamental
tests of quantum mechanics and as a basic resource of quantum information
processing [1]. Quantifying the amount of entanglement between quantum
systems is a recent pursuit that has attracted a diverse range of
researchers [5-15]. When we look at the entanglement of the mixed state as a
whole, can we still calculate the relative entropy of entanglement? This is
in general very difficult to do for multipartite mixed states, and some
partial methods for upper bounds have only been presented recently \textrm{%
[27]}. In this section, we will apply the results obtained previously to
derive the entanglement degree for a single multi-level atom interacting
with a cavity field without using the diagonal approximation method adapted
in [8,9]. With a certain unitary operator, the final state after the
interaction between the atom and the field is given by
\begin{eqnarray}
\mathcal{E}_{t}^{*}\rho &=&U_{t}\left( \rho \otimes \varpi \right) U_{t}^{*}
\nonumber \\
&=&\gamma _{1}U_{t}|a,\varpi \rangle \langle \varpi
,a|U_{t}^{*}+\sum\limits_{i=2}^{m-1}\gamma _{i}U_{t}|b_{i},\varpi \rangle
\langle \varpi ,b_{i}|U_{t}^{*}.
\end{eqnarray}
Therefore the von Neumann entropy of the total system is given by
\begin{equation}
S(\mathcal{E}_{t}^{*}\rho )=-\sum\limits_{i=1}^{m}\gamma _{i}\log \gamma
_{i}.  \label{entropy01}
\end{equation}
Taking the partial trace over the atomic system, we obtain
\[
\rho _{t}^{F}=tr_{A}\mathcal{E}_{t}^{*}\rho .
\]
Then the von Neumann entropy for the reduced state $S(\rho _{t}^{F})$ is
computed by
\begin{equation}
S(\rho _{t}^{F})=-\sum\limits_{i=1}^{m^{2}}\lambda _{i}^{F}(t)\log \lambda
_{i}^{F}(t),
\end{equation}
where $\{\lambda _{i}^{F}(t)\}$ are the solutions of
\begin{equation}
\det [\hat{\rho (t)}-\hat{\lambda (t)}\hat{N(t)}]=0,
\end{equation}
where $\hat{\rho (t)}$ and $\hat{N(t)}$ are $m^{2}\times m^{2}$ matrices
having the following elements
\begin{eqnarray}
\left[ \hat{\rho (t)}\right] _{ij} &\equiv &\langle \psi _{i}(t)|\rho
_{t}^{F}|\psi _{j}(t)\rangle ,\quad (i,j=1,2,3,...m^{2}),  \nonumber \\
\left[ \hat{N(t)}\right] _{ij} &\equiv &\langle \psi _{i}(t)|\psi
_{j}(t)\rangle ,\quad (i,j=1,2,3,...m^{2}),
\end{eqnarray}
and $|\psi _{j}(t)\rangle $ are the eigenfunctions of the following
eigenvalue problem $\rho _{t}^{F}|\psi _{i}(t)\rangle =\lambda
_{i}^{F}(t)|\psi _{i}(t)\rangle .$

On the other hand, the final state of the atomic system is given by taking
the partial trace over the field system:
\[
\rho _{t}^{A}\equiv tr_{F}\mathcal{E}_{t}^{*}\rho .
\]
Then the von Neumann entropy for the reduced state $S(\rho _{t}^{A})$ is
computed by
\begin{equation}
S(\rho _{t}^{A})=-\sum\limits_{i=1}^{m}\lambda _{i}^{A}(t)\log \lambda
_{i}^{A}(t),  \label{entropy03}
\end{equation}
where $\lambda _{i}^{A}(t)$ can be calculated by obtaining the eigenvalues
of the reduced atomic state. Using the above equations, the final expression
for the entanglement degree in the $m$-level system takes the following form
\begin{eqnarray}
I_{\mathcal{E}_{t}^{*}\rho }\left( \rho _{t}^{A},\rho _{t}^{F}\right)
&\equiv &tr\mathcal{E}_{t}^{*}\rho (\log \mathcal{E}_{t}^{*}\rho -\log (\rho
_{t}^{A}\otimes \rho _{t}^{F}))  \nonumber \\
&=&\sum\limits_{i=1}^{m}\gamma _{i}\log \gamma
_{i}-\sum\limits_{i=1}^{m^{2}}\lambda _{i}^{F}(t)\log \lambda
_{i}^{F}(t)-\sum\limits_{i=1}^{m}\lambda _{i}^{A}(t)\log \lambda _{i}^{A}(t).
\end{eqnarray}
\textrm{It turns out to be rather easy to derive an analytic expression for
the entanglement degree for any given system, since }with the help of
equation (23) it is possible to study the entanglement degree of any $m$%
-level system when the system starts from its mixed state.

\begin{figure}[tbph]
\begin{center}
\includegraphics[width=11cm]{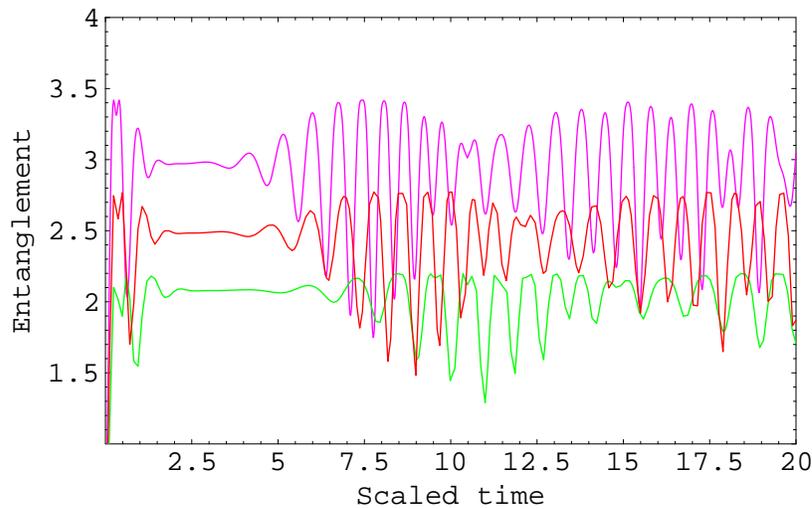}
\end{center}
\caption{The evolution of the entanglement degree $I_{\mathcal{E}%
_{t}^{*}\rho }\left( \rho _{t}^{A},\rho _{t}^{F}\right) $ as a function of
the scaled time. The mean photon number $\bar n=5$, and the detuning
parameter $\Delta $ has zero value, where, from bottom to top depicts
three-, four- and five-level atom, respectively. }
\end{figure}

\begin{figure}[tbph]
\begin{center}
\includegraphics[width=11cm]{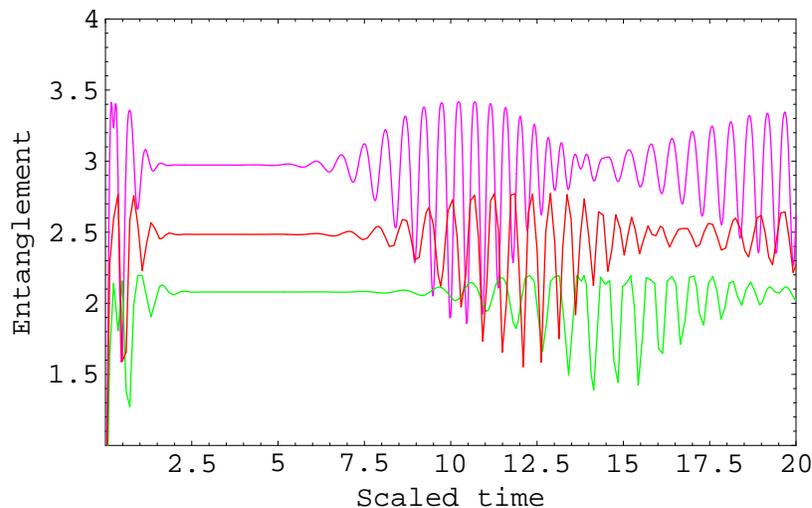}
\end{center}
\caption{The same as in figure 1 but now $\bar n=10$. }
\end{figure}

This seems significant, and one then wonders whether the trend might
continue with the general multi-atom (or ions) case. That is, whether one
might be able to consider more than one atom and still be able to measure
the entanglement degree using quasi-mutual entropy. Also, one might wonder
whether a similar effect could carry over to different field states. That
would be very nice because one could contemplate different protocols for
different initial states of the field. To go a step further towards a
deterministic entanglement degree, we note a peculiar effect in the present
paper: we get more entanglement with increasing $m$ (number of levels).
Indeed in the limit that $\gamma _{i}\sim 0,\quad i>1,$ (i.e. $\gamma
_{1}\approx 1),$ the entanglement degree is only twice of the quantum field
(atomic) entropy. In the general case ($i.e.,$ $\gamma _{1}\neq $ $1$), the
final state does not necessarily become a pure state, so that we need to
make use of $I_{\mathcal{E}_{t}^{*}\rho }\left( \rho _{t}^{A},\rho
_{t}^{F}\right) $ in order to measure the degree of entanglement in the
present model. Thus our initial setting enables us to discuss the variation
of the entanglement degree for different values of the parameter $\gamma
_{i} $ of the initial atomic system. \textrm{A related model allowing an
analytic treatment of the mixed-state entanglement as well as valuable
insight, namely the two-level atom (}$m=2)$ has been discussed in [8,9]%
\textrm{. An example of a truly mixed state for which the entanglement
manipulations have been proven to be asymptotically reversible has been
reported in Ref. [28]. }

\textrm{Here we focus on the time development of the entanglement degree for
some special cases such as three-, four- and five-level atoms. }In figure 1,
we plot the function $I_{\mathcal{E}_{t}^{*}\rho }\left( \rho _{t}^{A},\rho
_{t}^{F}\right) $ which describes the entanglement degree in the case when
the field is initially in a coherent state squeezed state with a mean photon
number $\overline{n}=5,$ and the mixed state parameters $\gamma _{1}=0.99$.
In this case we see that, the entanglement degree function oscillates around
values nearly equals the maximum values (2$\ln (m)$). Let us remark that, in
the pure state case, the von Neumann entropy is limited by $\ln (m)$, and
then $I_{\mathcal{E}_{t}^{*}\rho }\left( \rho _{t}^{A},\rho _{t}^{F}\right) $
reduces to 2$\ln (m)$. From this figure we can say that the maximum value of
entanglement degree $I_{\mathcal{E}_{t}^{*}\rho }\left( \rho _{t}^{A},\rho
_{t}^{F}\right) $ is increased as the number of levels is increased.

Nevertheless, the minimum values lie within the region between the
two maximum values occurring in a similar way for different number
of levels, such that with higher $m$ the minimum values of the
entanglement degree occur at earlier times. In fact, for some higher
values of $m$ there were no persisting periods found to lie between
the maximum and minimum values. These results strongly indicate that
the higher number of levels give higher entanglement as well as more
oscillations. Figure 2, indicates that when the mean photon number
is increased further the minimum values of the entanglement occur at
later times.

\begin{figure}[tbph]
\begin{center}
\includegraphics[width=11cm]{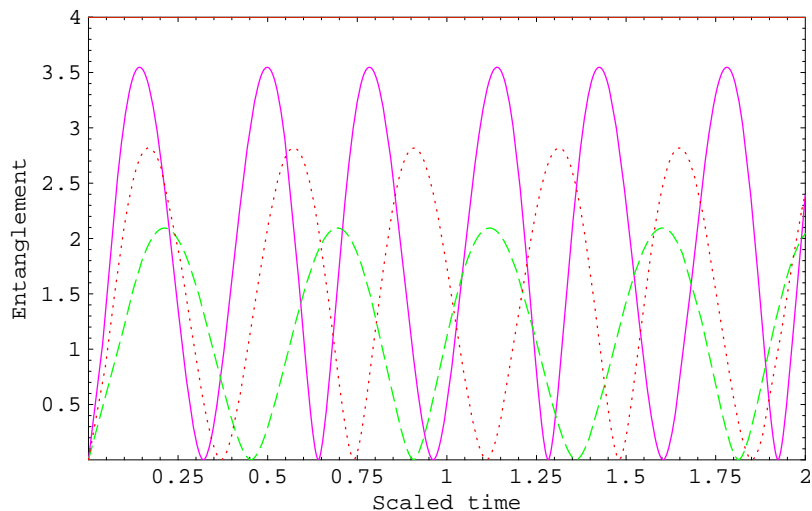}
\end{center}
\caption{The evolution of the entanglement degree $I_{\mathcal{E}%
_{t}^{*}\rho }\left( \rho _{t}^{A},\rho _{t}^{F}\right) $ as a function of
the scaled time. In this figure we consider the Fock state with $n=5$,
where, from bottom to top depicts three-, four- and five-level atom,
respectively. }
\end{figure}
In figure 3, we consider the entanglement degree as a function of the scaled
time with the field initially in a Fock state. The Fock state of the
electromagnetic field is very difficult to produce in experiments.
Nevertheless, these states are very important in quantum optics because of
their intrinsic quantum nature. This case is quite interesting because the
entanglement degree function oscillates around the maximum and minimum
values in time. We have shown here a new phenomena where the periodic
oscillations occur irrespective of number of atomic levels involved. This
reflects the various influences of the initial states of the field. A slight
change in $n$ therefore, dramatically alters the entanglement. It should be
noted that for a special choice of the initial state setting, the situation
becomes interesting where we find that a higher multi-level atom interacting
with an initially coherent field exhibits superstructures instead of the
usual first-order revivals.

It is worth mentioning that the dynamics of quantum multi-level systems has
always been of interest, but has recently attracted even more attention
because of application in quantum computing. Several systems have been
suggested as physical realizations of quantum bits allowing for the needed
control manipulations, and for some of them the first elementary steps have
been demonstrated in experiments [\textrm{29]}.

\section{Conclusion}

Summarizing, we have shown how to determine the maximum and minimum possible
values of the entanglement degree for multi-level atoms interacting with a
cavity field. The forms of states that achieve these maximum and minimum
values are the same as those for the case of the von Neumann entropy if we
consider the pure state case. These results are applicable for measuring the
entanglement for mixed states in any multi-level systems. We have identified
the relation between the entanglement measures that is necessary for these
mixed states setting. This relation holds between the quasi-mutual entropy
and von Neumann entropy. The general formula we have derived may carry over
to any multi-level system. For the examples we have examined the
entanglement degree for three-, four- and five-level atoms, we have found
that as the number of levels increases the maximum values of the
entanglement degree also increases, but these values are achieved for
earlier times when the number of levels is increased accordingly.

An open and very interesting question is whether the quasi-mutual entropy
technique that we have described here can be transferred to other systems in
which atomic and cavity decays are present. In those systems it may be
possible to augment or simplify the definition of the quasi-mutual entropy
making its applications more accessible.
\[
\]
{\Large Acknowledgments}

~

M. Abdel-Aty wishes to express his thanks for the financial support
and the hospitality extended to him at the IIU University Malaysia.
M. R. B. Wahiddin acknowledges the support of Malaysia IRPA research
grant 09-02-08-0203-EA002.

~

\end{document}